\documentstyle[12pt,epsfig]{article}
\title
{\bf Peculiarities of localization of several sonoluminescent
bubbles in spherical resonators}

\author{V.B. Belyaev$^1 $, B.F. Kostenko$^1 $, M.B. Miller$^2 $, and A.V. Sermyagin$^2 $}
\begin{document}
\maketitle
\begin{center}
$^1 $ Joint Institute for Nuclear Research, Dubna
\\ $^2 $ Institute in Physical--Technical Problems, Dubna
\end{center}

%\vspace{1cm}
\begin{center}
\bf Abstract
\end{center}

%\vspace{0.3cm}

%\begin{sloppypar}

Experiments on generation of 1, 2, 4, and 6 sonoluminescent
bubbles in water with an external ultrasound source in an acoustic
sphere resonator with glass walls have been carried out.
Theoretical examination has shown that the observed excitation
frequencies could be described with a good accuracy taking into
account that the velocities and pressures of the contacting media
on the external and internal resonator surfaces are equal. The
necessity of accounting for oscillations with non--zero
self--values of angular momentum operator has been shown when
describing the features of localization of several bubbles. To
explain a strangely small distance between the bubbles in the case
of two--bubble sonoluminescence the following possible
explanations have been proposed: a) mechanism of space splitting
of a mode with a singular angular momentum and b) mechanism of
secondary excitation when one of the bubbles is trapped into the
acoustic trap created by high--frequency vibrations arising
simultaneously when the other bubble sonoluminescence occurs.

%\end{sloppypar}

\eject

\section{Introduction}

$\qquad$The phenomenon of sonoluminescence -- light emission by
gas--vapor bubbles under the influence of acoustic cavitation --
was observed in 1930 during examination of chemical reactions
initiated by ultrasound. The great interest in this problem was
aroused from the early '90s due to discovery of the effect of
single bubble sonoluminescence in the field of a standing sound
wave \cite{Gaitan}. The most part of the experimental and
theoretical works on this subject are dealing with its different
aspects (see, for example, the reviews [2--4]). Less studied
remain the processes where several sonoluminescent bubbles are
involved, particularly, their space localization inside the
acoustic resonator is insufficiently studied. To describe standing
waves in a glass sphere resonator, it is proposed in
\cite{Barber1} to consider the velocity potential value on the
glass--air boundary equal to zero and the transition from liquid
to glass was neglected. The analysis of our experiments has shown,
however, that within this approximation it is impossible to
explain the observed resonance frequencies of the small number
(one to six) of sonoluminescent bubbles. In addition, in our
experiments when several sonoluminescent bubbles were excited,
they formed rather stable space structures, which cannot be easily
described with spherical symmetric velocity potentials
corresponding to zero orbital momentum (described in
\cite{Barber1}). This work describes the experiments on generation
of several sonoluminescent bubbles in a sphere resonator as well
as theoretical analysis of the results obtained.

\section{Experiments}
$\qquad$Present experiments were aimed at studying special
features of SBSL in acoustical resonators made of glass with
configurations slightly different from spherical.

\subsection{Resonators}

{\it Resonator $\#1$}. Spherical chemical retort with a long narrow
neck. The maximum size of the retort was 64.5 mm, minimum one 63.0
mm; wall thickness 1.0 mm. The neck: length 105.0 mm, external
diameter 7.5 mm, internal diameter 4.5 mm.\\{\it Resonator $\#2$.}
Spherical chemical retort with a spherical bottom and wider neck.
Diameter of the retort 63.0 mm, thickness of the wall 1.5 mm,
diameter of the neck 20 mm, length 30 mm.\\{\it Resonator $\#3$.} Thin
wall retort with ellipsoid--of--revolution form, and a short neck.
Maximum and minimum retort cross sections: 62.0 and 59.0 mm. Neck:
diameter 11 mm, length 5 mm.

During the experiments Resonators $\#1$ and $\#2$ were supported by their
necks with the help of laboratory three--fingered clamp stand.
Resonator $\#3$ was placed into a type of a string--bag made of strong
sewing suspended on a stand.

\subsection{Electroacoustic transducers.}

To induce the vibrations three types of hollow cylinder
transducers were used.\\ {\it Transducer $\#1$.} Diameters: 30$/$26
mm, height: 16 mm.\\ {\it Transducer $\#2$.} Diameters: 18.5$/$16.0
mm, height: 22 mm.\\ {\it Transducer $\#3$.} Diameters: 18.5$/$16.0
mm, height: 10 mm.

As a sensor of acoustical vibrations a self--made piezoelectric
cells were used. Resonator $\#1$ was equipped with one activator
located within the equatorial plane, Resonator $\#2$ was equipped with
two activators located symmetrical within the equatorial plane.
Resonator $\#3$ was equipped with one activator located at the bottom
of the retort.

\subsection{Electronics}
Activators were connected to the transformer output of low
frequency 100 W amplifier with 60 kHz  band pass and 150 V
maximum voltage. The amplifier was fed by low--frequency
generator G3--110. The scheme contains also series LC circuit:
L -- variable inductance of a coil with a ferrite core, C --
capacitance of activator; the circuit being tuned to one of
the fundamental frequencies of a resonator.

\subsection{Seeding of cavitation centers}
Seeding of cavitation centers was carried out by dropping a
special nail--headed rod onto a surface of water in the base of
resonator necks, or by injection of air bubbles into the water
by a microsyringe in the case of Resonator $\#3$.

\subsection{Results}
In the case of Resonator $\#1$ a satellite bubble was observed,
i.\,e. a second luminescent bubble. Brightness of this second
bubble was much less then that of the main bubble. Satellite
oscillated a little. Both bubbles located along the axis of the
neck of the retort above (the main bubble), and below (the
satellite) the center of the retort. The frequency was $f=27048$
kHz, the quality factor of the system $Q=1000$.\\In Resonator
$\#2$ there were four luminescent bubbles in the vertexes of an
imaginary quadrant. Plane of the quadrant changed a position a
little depending on the frequency.\\In Resonator $\#3$, a stable
luminescence took place concurrently for six bubbles, alternately
fading and arising again in different positions. The frequency was
$f=53840$ kHz (the second harmonic), the quality factor $Q=234$.

\section{Excitation of single sonoluminescent   \\ bubbles}

$\qquad$It is known that standing monochromatic acoustic waves of
spherical form are described with solutions
\begin{equation}\label{1}
\psi  = Ae^{ - i\omega t} \frac{{\sin kr}} {r}, \qquad k =
\frac{\omega } {c}
\end{equation}
of radial wave equation
\[
\frac{1} {{c^2 }}\frac{{\partial ^2 \psi }} {{\partial t^2 }} =
\frac{1} {{r^2 }}\frac{\partial } {{\partial r}}(r^2
\frac{{\partial \psi }} {{\partial r}})
\]
for the velocity potential  $\psi$. This solutions satisfy the
finiteness condition in a point $r=0$, which is true if there is
no source at the point of origin  \cite{Landau1}. We will assume
for liquid contained in the spherical flask that the acoustic
field inside it is described by a potential (\ref{1}), and inside
its wall with a potential
\begin{equation}\label{2}
 \psi'  = A'e^{ - i\omega t} \frac{{\sin (k\;'r + \alpha )}} {r},
\qquad k\;' = k\frac{c} {{c\;'} ,}
\end{equation}
where $c$ and $c'$ are sound velocity in water and in glass,
respectively, and $\alpha$ is a phase shift caused by the
transition of the wave to the material with different mechanical
properties. The condition that the pressure on the outer side of
the wall is equal to zero (we neglect wave radiation into
environment) gives
\[
p|_{r = a_ +  }  =  - \rho '\frac{{\partial \psi '}} {{\partial
t}}|_{r = a_ +  }  = 0,
\]
where  $a_+ \equiv a+d$ is an external radius of the flask and $d$
is the wall thickness. From here we have
\[
  k\;'a_ +   + \alpha  = n\pi , \qquad n = 0,1,2,...
\]
Further we will assume that $n=1$ (it appeared that this very $n$
value agrees with the experiment), which corresponds to the
solution
\begin{equation}\label{3}
\psi'  =  - A'e^{ - i\omega t} \frac{{\sin k\;'(r - a_ +  )}} {r}.
\end{equation}

We will describe wave reflection on the water--glass boundary
accounting for the requirement of medium continuity and the
condition of pressure equality at the both sides of the boundary
\cite{Skudrzyk}. The last condition gives
\[
\rho A\frac{{\sin ka}} {a} = \rho 'A'\frac{{\sin k\;'d}} {a},
\]
or
\begin{equation}\label{4}
 A' = \frac{\rho } {{\rho \;'}}\frac{{\sin ka}} {{\sin k\;'d}}A.
\end{equation}
The requirement of medium continuity on the surface $r=a$ is
reduced to the equality in the points of velocity of both media $
v
=
\partial \psi / \partial r $ and $ v' =
\partial \psi ' / \partial r $. Therefore, we find
 \[
A\left( {\frac{{k\cos ka}} {a} - \frac{{\sin ka}} {{a^2 }}}
\right) = A'\left( {\frac{{k'\cos k'd}} {a} + \frac{{\sin k'd}}
{{a^2 }}} \right).
\]
This with the account of  (\ref{4}), gives the following
transcendental equation for finding the wave number $k$:
\begin{equation}\label{5}
ka \; {\rm ctg} (ka) - 1 = \frac{\rho } {{\rho \;'}}\left[ {1 +
\frac{c} {{c\;'}} ka \; {\rm ctg} \left( {\frac{c} {{c\;'}}kd}
\right)} \right] ,
\end{equation}
where $c/c\;'\simeq 0.25 \div 0.3$ and $\rho / \rho \;' \simeq 0.4
$
are the ratios of sound velocities and material densities of water
and glass. First solution  (\ref{5}) with a minimum value of wave
number $k$ corresponds to  $ka \simeq 3.685$, which, for example,
for $a=3.15$ corresponds to the glass thickness $d \simeq 2.6$ mm,
which is rather close to a real one. Acoustic oscillation
frequency is connected with the wave vector $k$ by the relation
$f=c k/2\pi$, which also gives a value close to the observed $f
\simeq 27$ KHz.

According to the existing theory which is in a good agreement with
the experiment, cavitation bubbles are located, depending on their
size, either in nodes, or in points of local maximum of pressure
amplitude in standing acoustic wave (anti--nodes). Thus, small--size
bubbles (smaller than a certain resonance value) are located in
nodes, while big ones -- in points of maximum \cite{Yosioka}.
Experiments with several sonoluminescent bubbles have shown that
in this sense they are ``big" since they are located in anti--nodes
\cite{Barber1}.

Equation  (\ref{5}) allows one to identify the experimentally
observed bubble trapping at a center of water--filled glass flasks
with outer radiuses $a_+ $, equal to 3.15 and 3.20 cm, at
excitation frequencies 27 and 26.5 KHz, respectively, as their
trapping by the acoustic traps corresponding to the solutions
(\ref{5}) with minimum values of wave number  $k$ i.e., with a
first harmonics of the resonator self--oscillations). At the same
time, the bubble observed at the frequency of 53.82 KHz at the
center of the water--filled glass flask at $a_+ = 3.025 \pm 0.075 $
cm, can be corresponded with the second harmonics. Single
sonoluminescent bubbles corresponding to third and higher
harmonics have not been studied in these experiments.

\section{ Excitation of two, four, and six  \\ sonoluminescent bubbles}

$\qquad$Since at acoustic wave excitation under real conditions
the spherical symmetry condition is violated surface
excitations in the flask should be described with the following
function:

\[
r_s (\theta ,\varphi ,t) = R(t) + \sum\limits_{l,m} {a_{lm} }
(t)Y_{lm} (\theta ,\varphi ),
\]
where  $Y_{lm}$ are spherical functions accounting for the
dependence of surface deformations on $\theta $ and $\varphi$
angular coordinates.
Then the wave equation for the potential
should be as well taken in the general form
\[
\frac{1} {{c^2 }}\frac{{\partial ^2 \psi }} {{\partial t^2 }} =
\frac{1} {{r^2 }}\frac{\partial } {{\partial r}}\left( {r^2
\frac{{\partial \psi }} {{\partial r}}} \right) - \frac{{\hat l^2
}} {{r^2 }}\psi ,
\]
where
\[
\hat l^2  =  - \left[ {\frac{1} {{\sin \theta }}\frac{\partial }
{{\partial \theta }}\sin \theta \frac{\partial } {{\partial \theta
}} + \frac{1} {{\sin ^2 \theta }}\frac{{\partial ^2 }} {{\partial
\varphi ^2 }}} \right]
\]
is the ``angular momentum"\footnote{Since angular variables in
equations (4.1) and (4.2) describe anisotropy and not rotary
motion, the notion ``angular momentum" should be understood here
as a certain convention suitable for classification of the
solutions.} , which satisfies the condition
\[
\hat l^2 Y_{lm}  = l(l + 1)Y_{lm}.
\]
For the monochromatic wave with the potential
\[
\psi  = e^{ - i\omega t} R(r)Y_{lm} (\theta ,\varphi )
\]
The radical wave equation takes the form of
\begin{equation}\label{6}
\frac{1} {{r^2 }}\frac{\partial } {{\partial r}}\left( {r^2
\frac{{\partial R}} {{\partial r}}} \right) + \left( {k^2  -
\frac{{l(l + 1)}} {{r^2 }}} \right)R = 0,
\end{equation}
%где, как и прежде, $k = \omega /c.$
from where it can be seen that the radical part of the potential
depends on $k$ and $l$ parameters and does not depend of $m$:
\[
R = R_{kl} (r).
\]

The solution of equation (\ref{6}), satisfying the finality
condition in the point $r=0$, has the form
\[
R_{k0} (r) = \frac{{\sin kr}} {r},
\]
in agreement with the results of the previous section. General
solution of equation (\ref{6}), satisfying the finality condition
in the point  $r=0$, has the form (see, for example,
\cite{Landau2})
\begin{equation}\label{7}
R_{kl} (r) = Cr^l \left( {\frac{1} {r}\frac{d} {{dr}}} \right)^l
\frac{{\sin kr}} {r} = cj_e (kr),
\end{equation}

where $j_e (k)$ is the spherical Bessel function.
Particularly,
\[
  R_{k1} (r) = C_1 \left[ {k\frac{{\cos kr}}
{r} - \frac{{\sin kr}} {{r^2 }}} \right],
\]
\[
  R_{k2} (r) = C_2 \left[ {\left( {\frac{3}
{{r^3 }} - \frac{{k^2 }} {r}} \right)\sin kr - \frac{{3k\cos r}}
{{r^2 }}} \right] \hfill \\
\]
etc. The velocity potential
\begin{equation}\label{8}
\psi _{klm}  = Ce^{ - i\omega t} R_{kl} (r)Y_{lm} (\theta ,\varphi
)
\end{equation}
corresponds to several oscillating out of phase standing spherical
waves the symmetry centers of which are very close to each other.
For example, we have for two centers

\begin{equation}\label{9}
\psi = C\left(\frac{\sin kr_2 }{r_2} - \frac{\sin kr_1 }{r_1 }
\right) = C\frac{\partial } {{\partial r}}\left( \frac{\sin kr
}{r} \right) \frac{\partial r}{\partial s} \; ds = C^{\;'}
\frac{\partial } {{\partial r}}\left( \frac{\sin kr}{r}
\right)\cos \theta ,
\end{equation}
where $r_1$  and $r_2$ are the distances from the centers to the
observation point, $ds$ is the distance between the centers, which
is assumed to be infinitely small, $\partial r /ds = \cos \theta$
is a direction cosine of the observation point and the line
connecting the centers. This potential form corresponds to  $l=1$.
Similarly the velocity potentials, corresponding to the higher
angular momentums, are interpreted.

The points of pressure extremals location showing bubble position
in the resonator, are described with a system of equations

\[
\qquad \qquad \qquad \qquad \qquad \qquad \left\{
\begin{array}{l}
{\large d\psi  /dr = 0},  \\
{\large d\psi  /d \theta  = 0}, \qquad \qquad \qquad \qquad \qquad \qquad \qquad \qquad (10) \\
{\large d\psi  /d \varphi  = 0}.
 \end{array} \right.
\]
The velocity potential  $\psi$ at a preset excitation frequency
$f_{kl}$  is, generally speaking, a superposition of functions
(\ref{8}) with different values of $m$ projections of angular
momentum, $m=-l, ..., +l$. In the case of frequency degeneration
(within the $\Delta f$ width determined by the system $q$--factor
by $l$ parameter , it is necessary as well to take into account
the superpositions of the states with different values of orbital
momentum.  For separate $\psi_{klm}$ modes the system (10) can be
simplified and obtains the following form:
\[
\qquad \qquad \qquad \qquad \qquad \qquad \left\{
\begin{array}{l}
{\large d R_{kl}  /dr = 0},  \\
{\large d Y_{lm}  /d \theta  = 0}, \; \; \; \; \qquad \qquad \qquad \qquad \qquad \qquad \qquad (11) \\
{\large d Y_{lm}   /d \varphi  = 0}.
 \end{array} \right.
\]

From  (11) for the $Y_{1,\pm 1}$ mode corresponding to $l=1$, $$
\theta_1 =0, \qquad \theta_2 =0, $$ $$ \varphi_1=0, \qquad
\varphi_2=0. $$ The obtained angular coordinates correspond to the
experimentally observed at the frequency $f\simeq 49.5$ KHz
four--bubble configuration located in the flask with the radius
$a_+ = 3.2$~см at the vertexes of an imaginary square.

If for the approximate estimate of probable distances from the
bubbles to the center of the flask according to  \cite{Barber1},
we take soft conditions at the external boundary
\[
p|_{r = a_ +  }  =  - \rho \frac{{\partial \psi }} {{\partial
t}}|_{r = a_ +  }  = 0,
\]
where   $\psi$ is the velocity potential of acoustic oscillations
in liquid, then, accounting for the explicit expression for
$R_{k1}$, we obtain the equation $$ tg \; ka_+ = ka_+ , $$ the
solutions of which are $$ ka_+ \simeq \qquad 4.4934, \qquad
7.7253, \qquad 10.9041 \qquad {\rm etc.\;} $$ Taking into account
that $k= 2\pi f/c \simeq 2.075$ cm$^{-1}$, we find that only the
first of these solutions corresponds to the experimentally
observed excitation frequency. Probable positions of the bubbles
$r_i$, are determined from the condition $$ \frac{\partial
p}{\partial r} =0, $$ which leads to the equation $$ tg\; kr_i =
\frac{2kr_i }{2 - (kr_i )^2}. $$ It can be seen that the first of
the solutions $$ kr_i \simeq \qquad 4.4934, \qquad 7.7253 \qquad
\rm etc.\; , $$ of this equation with the account of $k$ value
gives $$ r_1 \simeq 2.2\; cm, $$ which is somewhat greater than
the experimentally observed value $r_{exp} \simeq 1.5$ cm. This
shows that the speculations of the previous section, accounting
not only for the external boundary conditions but for the
water--glass interface conditions as well, should be true for the
case of various sonoluminescent bubbles (for each of the radial
functions $R_{kl}$).

It would be natural to take as a potential a combination of states
corresponding to $l=2$ for the six--bubble configuration observed
at a frequency of $f\simeq 53.8$~KHz located in the flask at the
same plane approximately one centimeter from the glass wall. In
this case the solution of system (11) for angular coordinates
gives
\[
\theta  = \frac{\pi } {4},\; \frac{{3\pi }} {4}; \qquad  \varphi
= 0,\pi
\]
for  $Y_{2, \pm 1}$;
\[
  \theta  = 0,\; \frac{\pi }
{2},\; \pi ;\qquad   \varphi  = 0,\; \frac{\pi } {2},\; \pi
,\frac{{3\pi }} {2}
\]

for $Y_{2, \pm 2}$. On the assumption that not all of the bubble
traps were filled up (though no special measures had been
undertaken for this purpose), among the solutions obtained one
could find the sets similar to the experimentally observed
configuration. It is especially interesting to analyze a
two-bubble configuration located along the vertical line near the
center of the flask (with a radius $a_+ = 3.15$ cm) at $f\simeq
26.9603$ KHz, frequency where one of the bubbles was emitting
light more intensely than the other. The distance between them is
unexpectedly small (only few millimeters), which is difficult to
explain with the use of equation (11) for a radical part of the
potential\footnote{Small distances of bubbles from the center of
the flask correspond to large values of the wave number. In this
case, however, $k=1.13$ cm$^{-1}$.}. One of the possible
explanations of this fact is that in this case one of the bubbles
was trapped by an acoustic trap created by the high--frequency
pulsations of the other arising simultaneously with its
sonoluminescence. It is well known that the wave length of these
high--frequency oscillations is approximately 20 times smaller than
the standing wave length exciting sonoluminescence \cite{Barber1},
and in our case is within the millimeter range. If the flask had
the shape of a spheroid with a small distance between the focuses
and if one of the bubbles was located at one of the focuses, then
the emitted by it high--frequency waves should be accumulated at
the other focus \cite{geometry} thus creating an acoustic trap
capable of supporting sonoluminescence of another bubble. The
other possible explanation of these results is connected with
formula (9), which shows that the solutions corresponding to
$l=1$, and the configuration of two bubbles, located at a finite
though small distance between each other, are close. The
transition of the common solution with $l=1$ into the
configuration of two bubbles oscillating in anti--phase is possible
on the assumption of a certain instability connected with
excitation of vertical oscillations due either to the fact that
the flask has a neck, or to periodic deformations of the surface
of the flask by the attached piezoelectric transducer .
At present this problem remains open.

\section{Conclusion}
$\qquad$This paper describes an experimental setup permitting to
study the processes of single-- or several--bubble sonoluminescence
in a sphere resonator. Though seemingly simple, an experiment of
this kind is not a trivial task and it requires a number of
conditions (such as fine tuning of the excitation frequency,
requirements to the resonator $Q$--factor, temperature, and pressure
in cavitating liquid, etc.). At the same time it is very important
to study these processes if regarded that at present no generally
accepted theory exists explaining the nature of the observed light
emission  \cite{Barber1,Margulis1}, and also because recently the
chances appeared to attain extremely high temperature and
pressures in these processes \cite{Barber1,Taleyarkhan,Belyaev}.
Therefore, it is clear that the precise description of the
conditions under which stable laboratory reproducibility of these
phenomena is possible is of much importance.

The analysis performed in this paper has shown that when
calculating resonant excitation frequencies, at which one could
expect sonoluminescent bubble arising, boundary conditions should
be accounted for not only at the external but at the internal side
of the resonator wall as well. In the case of several bubbles
excitation, radial oscillation modes with different from zero
orbital momentum should be accounted for. Further, to check the
correctness of the proposed description one should have the
precise measurements of bubble position in the resonator with the
account of optical distortions due to light passing through the
water--filled glass flask. It seems important as well to study the
reason of the anomalous approach of two sonoluminescent bubbles.

This work was supported by the International Scientific--Technical
Center (Project No. 1471) and by the Russian Foundation for Basic Research
((Projects Nos. 02--01--00606 and 02--02--16397).

\vspace{1cm}

%\eject
%\newpage


\begin{thebibliography}{10}

\bibitem{Gaitan}  {\it Gaitan  D.F.,  Crum L.A.,  Roy R.A.}// J. Acoust. Soc.
Am. 1992. V. 91. PP.~3166--3172.

\bibitem{Barber1} {\it Barber  B.P.,  Hiller R.A., Lofstedt R.,  Putterman S.J.,
 Weninger K.R.} // Phys. Rep. 1997. V.281. PP.~65--143.

\bibitem{Putterman1} {\it Putterman S.J.,  Weninger K.R.} // Annu. Rev. Fluid.
Mech. 2000. V.32. PP.~445--476.

\bibitem{Margulis1} {\it Margulis M.A.}// UFN. 2000. V.170. Pp.~263--287.

\bibitem{Landau1} ] Landau L.D., Lifshits E.M. Hydrodyunamics. M., Nauka. 1986. P. ~379.

\bibitem{Skudrzyk}  Skudrzyk E. Bases of acoustics. Volume I. M. Edit. IL, 1958. P. ~141.

\bibitem{Yosioka} Yosioka  K.,  Kawasima Y., Hirano H.  Acustica.
1955. Vol.5.  PP.~173 -- 178.

\bibitem{Landau2}  ] Landau L.D., Lifshits E.M. Quantum Dynamics. M., Nauka. 1974. P. ~135.

\bibitem{Taleyarkhan}  Taleyarkhan R.P., West C.D.,  Cho J.S., Lahey Jr. R.T.,
Nigmatulin  R.I.,  Block R.C.  Science. 2002. V.295.
PP.1868--1873.

\bibitem{Belyaev} Beliaev V.B., Kostenko B.F., Miller M.B., Sermyagin A.V.,
and Topolnikov A.S. Ultrahigh temperatures and acoustic
cavitation. 2003. JINR Communication P3--2003--214. Dubna.

\bibitem{geometry} Encyclopedia of elementary mathematics, volume 5. M. Nauka, 1966. P. ~575.

\end{thebibliography}
\end{document}